\documentstyle[12pt,graphicx,epsf]{article}
\begin{document}

\def\AEF{A.E. Faraggi}
\def\NPB#1#2#3{{\it Nucl.\ Phys.}\/ {\bf B#1} (#2) #3}
\def\PLB#1#2#3{{\it Phys.\ Lett.}\/ {\bf B#1} (#2) #3}
\def\PRD#1#2#3{{\it Phys.\ Rev.}\/ {\bf D#1} (#2) #3}
\def\PRL#1#2#3{{\it Phys.\ Rev.\ Lett.}\/ {\bf #1} (#2) #3}
\def\PRT#1#2#3{{\it Phys.\ Rep.}\/ {\bf#1} (#2) #3}
\def\MODA#1#2#3{{\it Mod.\ Phys.\ Lett.}\/ {\bf A#1} (#2) #3}
\def\IJMP#1#2#3{{\it Int.\ J.\ Mod.\ Phys.}\/ {\bf A#1} (#2) #3}
\def\nuvc#1#2#3{{\it Nuovo Cimento}\/ {\bf #1A} (#2) #3}
\def\RPP#1#2#3{{\it Rept.\ Prog.\ Phys.}\/ {\bf #1} (#2) #3}
\def\APJ#1#2#3{{\it Astrophys.\ J.}\/ {\bf #1} (#2) #3}
\def\APP#1#2#3{{\it Astropart.\ Phys.}\/ {\bf #1} (#2) #3}
\def\etal{{\it et al\/}}

\newcommand{\bev}{\begin{verbatim}}
\newcommand{\beq}{\begin{equation}}
\newcommand{\beqa}{\begin{eqnarray}}
\newcommand{\beqn}{\begin{eqnarray}}
\newcommand{\eeqn}{\end{eqnarray}}
\newcommand{\eeqa}{\end{eqnarray}}
\newcommand{\eeq}{\end{equation}}
\newcommand{\Eev}{\end{verbatim}}
\newcommand{\bec}{\begin{center}}
\newcommand{\eec}{\end{center}}
\def\ie{{\it i.e.}}
\def\eg{{\it e.g.}}
\def\half{{\textstyle{1\over 2}}}
\def\nicefrac#1#2{\hbox{${#1\over #2}$}}
\def\third{{\textstyle {1\over3}}}
\def\quarter{{\textstyle {1\over4}}}
\def\m{{\tt -}}
\def\mass{M_{l^+ l^-}}
\def\p{{\tt +}}

\def\slash#1{#1\hskip-6pt/\hskip6pt}
\def\slk{\slash{k}}
\def\GeV{\;{\rm GeV}}
\def\TeV{\;{\rm TeV}}
\def\y{\;{\rm y}}

\def\l{\langle}
\def\r{\rangle}
\newcommand{\lsim}   {\mathrel{\mathop{\kern 0pt \rlap
  {\raise.2ex\hbox{$<$}}}
  \lower.9ex\hbox{\kern-.190em $\sim$}}}
\newcommand{\gsim}   {\mathrel{\mathop{\kern 0pt \rlap
  {\raise.2ex\hbox{$>$}}}
  \lower.9ex\hbox{\kern-.190em $\sim$}}}
\renewcommand{\thefootnote}{\fnsymbol{footnote}}
\setcounter{footnote}{0}

\begin{titlepage}
\samepage{
\setcounter{page}{1}

\rightline{OUTP-01-35P}
\rightline{UNILE-CBR-2001-4}
\rightline{July 2001}
\vspace{0.5cm}
\begin{center}
 {\Large \bf Stable Superstring Relics\\  
        and \\
        Ultrahigh Energy Cosmic Rays \\}
\vspace{0.5cm}
 {\large Claudio Corian\`{o}$^1$\footnote{Claudio.Coriano@le.infn.it},
           Alon E. Faraggi$^{2,3}$\footnote{faraggi@thphys.ox.ac.uk}
 and          Michael Pl\"umacher$^2$\footnote{pluemi@thphys.ox.ac.uk}\\
\vspace{.25cm}
{\it $^1$Dipartimento di Fisica,
 Universita' di Lecce,\\
 I.N.F.N. Sezione di Lecce,
Via Arnesano, 73100 Lecce, Italy\\}
\vspace{.25cm}
{\it $^2$Theoretical Physics Department,\\
University of Oxford, Oxford, OX1 3NP, United Kingdom\\}
\vspace{.25cm}
{\it $^3$ Theory Division, CERN, CH--1211 Geneva, Switzerland} }
\end{center}
\begin{abstract}
One of the most intriguing experimental results of recent years 
is the observation of Ultrahigh Energy Cosmic Rays (UHECRs) above
the GZK cutoff. 
Plausible candidates for the UHECR primaries are 
the decay products of a meta--stable matter state
with mass of order $O(10^{12-15}\;{\rm GeV})$, which simultaneously
is a good cold dark matter candidate.
We study
possible meta--stable matter states that arise from Wilson line breaking
of GUT symmetries in semi-realistic
heterotic string models.
In the models that we
study the exotic matter states can be classified according to patterns
of $SO(10)$ symmetry breaking. 
We show that cryptons, which are
states that carry fractional electric charge $\pm1/2$, 
and are confined by a hidden gauge group cannot produce
viable dark matter. This is due to the fact that, in addition to the
lightest neutral bound state, cryptons give
rise to meta--stable charged bound states.
However, these states may still account for the UHECR events. 
We argue that the uniton, which is an exotic Standard Model
quark but carries ``fractional'' $U(1)_{Z^\prime}$ charge,
as well as the singleton, which is a Standard Model singlet
with ``fractional'' $U(1)_{Z^\prime}$ charge, do provide
viable dark matter candidates and can at the same time explain the
observed UHECR events.

\end{abstract}
\smallskip}
\end{titlepage}

\section{Introduction}

One of the interesting experimental observations of recent
years is the detection of Ultrahigh Energy Cosmic Rays \cite{uhecr},
whose observed energy exceed the Greisen--Zatsepin--Kuzmin (GZK)
cutoff \cite{gzk}. There are apparently no astrophysical sources
in the local neighbourhood that can account for the 
events. The shower profile of the 
highest energy events is consistent with identification of the 
primary particle as a hadron but not as a photon or a neutrino.
The ultrahigh energy events observed in the air shower arrays
have muonic composition indicative of hadrons. 
The problem, however, is that the propagation of hadrons
over astrophysical distances is affected by the 
existence of the cosmic background radiation, resulting 
in the GZK cutoff on the maximum energy of cosmic ray 
nucleons $E_{\rm GZK}\le10^{20}\;{\rm eV}$. 
Similarly, photons of such high energies have a mean free path of less than
10Mpc due to scattering {}from the cosmic background radiation and 
radio photons. Thus, unless the primary is a neutrino, 
the sources must be nearby. On the other hand, the primary 
cannot be a neutrino because the neutrino interacts very weakly 
in the atmosphere. A neutrino primary would imply that the
depths of first scattering would be uniformly distributed 
in column density, which is contrary to the observations.

The difficulty 
in finding conventional explanations for UHECR opens the door for
innovative approaches. One of the most elegant possibilities
is that the UHECR originate {}from the decay of long--lived
super--heavy relics, with mass of the order of $10^{12-15}\;{\rm GeV}$
\cite{Berezinsky}.
In this case the primaries for the observed UHECR would originate
from decays in our galactic halo, and the GZK bound would not apply. 
Furthermore, the profile of the primary UHECR indicates that
the heavy particle should decay into electrically charged
or strongly interacting particles.

{}From the particle physics perspective two questions are of interest.
The first is the stabilization mechanism which produces
a super--heavy state with a lifetime of the order of the universe,
while still allowing it to decay and account for the observed UHECR events.
The second is how the required mass scale can arise naturally. In a
field theoretic model addressing both of these questions amounts to
fixing the required parameters by hand. It is therefore of further
interest to study the plausible states that may arise {}from string theory,
and whether such states can provide candidates for the UHECR events. 

One particular string theory candidate that has been 
proposed previously is the `crypton' \cite{eln,ben,sb}, in the context
of the flipped $SU(5)$ free fermionic string model \cite{revamp}. 
The `crypton' is a state that carries fractional 
electric charge $\pm1/2$ and transforms under a non--Abelian
hidden gauge group, which in the case of the flipped 
$SU(5)$ ``revamped''string model is $SU(4)$. The fractionally charged
states are confined and produce integrally charged hadrons
of the hidden sector gauge group. The lightest hidden hadron
is expected to be neutral with the heavier modes split
due to their electromagnetic interactions. A priori, therefore,
the `crypton' is an appealing CDM candidate, which is meta--stable
because of the 
fractional electric charge of the constituent `quarks'.
This implies that the decay of the exotic hadrons
can be generated only by highly suppressed non--renormalizable
operators. Effectively, therefore, the events that generate
the UHECR are produced by annihilation of the `cryptons'
in the confining hidden hadrons. Moreover, the mass scale
of the hidden hadrons is fixed by the hidden sector gauge
dynamics. Therefore, in the same way that the colour $SU(3)_C$
hadronic dynamics are fixed by the boundary conditions at
the Planck scale and the $SU(3)_C$ matter content, the 
hidden hadron dynamics are set by the same initial conditions
and by the hidden sector matter content.

However, we argue here that the `crypton' cannot in fact provide an appealing
candidate for CDM. The reason is again its fractional
electric charge. In addition to producing an electrically neutral hadron,
which is expected to be the lightest hidden hadron, integrally charged
hadrons are produced as well. The same reasoning that explains the
meta--stability of the lightest neutral hadron also implies that,
generically, the hidden, electrically charged, hadrons are semi--stable as
well. Namely, the constituent cryptons carry fractional electric
charge $\pm1/2$. Suppose we have an integrally charged tetron which
is composed of three $+1/2$, and one $-1/2$, constituents. In order
to convert the charged tetron to a neutral one, a $+1/2$ constituent
has to convert into a $-1/2$. However, since the cryptons are singlets
of $SU(2)_L$ this transition can only proceed via heavy GUT, or string,
modes which carry electric charge $+1$. Therefore, the transition {}from the
charged hadrons to the lightest neutral hadron can only proceed
by operators which are suppressed by the GUT, or string, unification
scales. Effectively, similar to the decay of the lightest
neutral hadron, this transition can be generated only by non--renormalizable
operators. Although one cannot rule out the possibility that in specific
models the decay of the neutral hadrons will arise {}from operators that
are suppressed relative to those that induce the charged hidden hadron
decays, generically, we expect both lifetimes to be of the same order
of magnitude. Therefore, in addition to the lightest neutral hadron
that could account for the UHECR, the `crypton' also gives rise
to long--lived charged hadrons, whose number densities are severely 
constrained. We also study the effect of intermediate scale cryptons
on the renormalization group running of the Standard Model gauge 
couplings and show that coupling unification necessitates the
existence of additional colour and electroweak states.

These arguments therefore prompt us to examine whether realistic 
string models may still produce well motivated candidates to account 
for the UHECR. We argue that the answer is affirmative. 
We study the different types of exotic
states that arise in semi--realistic string models and
their viability as candidates for producing the UHECR. 
We first discuss the classification of the various states in the 
string models and their properties. There are several
distinct categories of string states that may produce semi--stable
matter and we elaborate on the various cases. One general distinction
in the string models is between matter states that arise {}from the 
untwisted and twisted orbifold sectors, which 
do not break the GUT symmetry, versus those which arise {}from the
Wilsonian sectors, and which do break the GUT symmetry.
The former sector gives rise to Standard Model states which
preserve the GUT structure, whereas states from the Wilsonian sector
do not fit into multiplets of the original GUT symmetry and 
are meta--stable due to discrete symmetries.
Another category of states
are hidden sector glueballs that were suggested as candidates
for Self--Interacting Dark Matter (SIDM) \cite{maxim}.
The viable states, that may give rise to the UHECR events,
should possess two important properties.
First, they should transform under a non--Abelian hidden sector 
gauge group that confines at $O(10^{11-13})\;{\rm GeV}$.
Second, we argue that the most appealing candidates should be
Standard Model singlets. Such states can arise {}from the 
``Wilsonian'' sector and carry a non--$SO(10)$ charge
under the $U(1)_{Z^\prime}$ which is embedded in $SO(10)$, 
or they can arise {}from hidden sector glueballs. 

\setcounter{footnote}{0}
\section{Quasi--stable matter in realistic string models}

In this section we discuss the different classes of exotic states
in the string models that may produce UHECR candidates. For
concreteness we study these questions in the context of the 
realistic free fermionic heterotic string models.
A general remark on the different string theory constructions 
is that the heterotic string allows for the embedding of the 
Standard Model states in $SO(10)$ multiplets, whereas type I constructions
only permit constructions that have the generic structure
which is a product of $U(n)$ groups. The heterotic string
is therefore the only perturbative string theory which is compatible
with the orthodox unification scenario, which is highly motivated 
by the Standard Model multiplet structure and the MSSM gauge coupling
unification \cite{ross}. 

Heterotic string models generically give rise to exotic matter
states which arise because of the breaking of the non--Abelian unifying
gauge symmetry, $G$, by Wilson--lines \cite{yut,ww,ccf}.
The breaking of the gauge
symmetries by Wilson lines results in massless states that
do not fit into multiplets of the original unbroken
gauge symmetry. 
This is an important
property as it may result in conserved quantum numbers that will
indicate the stability of these so-called ``Wilsonian'' states.
The simplest example of this phenomenon is the existence
of states with fractional electric charge in the massless
spectrum of superstring models \cite{ww,eln,fcp}.
Such states are stable due to electric charge conservation.
As there exist strong constraints on their masses and abundance,
states with fractional electric charge must be diluted away or
be extremely massive. The same ``Wilson line''
breaking mechanism, which produces matter with fractional electric
charge, is also responsible for the existence of states which
carry the ``standard'' charges under the Standard Model gauge
group but which carry fractional charges under a different subgroup
of the unifying gauge group. For example, if the group $G$ is
$SO(10)$ then the ``Wilsonian'' states may carry non--standard
charges under the $U(1)_{Z^\prime}$ symmetry, which is embedded
in $SO(10)$ and is orthogonal to $U(1)_Y$.
Such states can therefore be long--lived if the $U(1)_{Z^\prime}$ gauge
symmetry remains unbroken down to low energies, or if some
residual local discrete symmetry is left unbroken after 
$U(1)_{Z^\prime}$ symmetry breaking.
What is noted is that the heterotic--string construction itself,
generically embodies the mechanism that results in semi--stable
heavy matter, which in turn may produce viable candidates for the
UHECR events. The existence of heavy stable ``Wilsonian''
matter can therefore be argued to be a ``smoking gun'' of 
heterotic string unification.

The realistic models in the free fermionic formulation are generated by
a basis of boundary condition vectors for all world--sheet fermions
\cite{revamp,fny,ALR,eu,SLM,cfn,lrs}.
The basis is constructed in two stages. The first stage consists
of the NAHE set, $\{{{\bf 1},S,b_1,b_2,b_3}\}$ \cite{nahe}. 
At the level of the NAHE set the gauge group is
$SO(10)\times SO(6)^3\times E_8$, with 48 generations and $N=1$
supersymmetry.
The NAHE set correspond to $Z_2\times Z_2$ orbifold compactification
with non--trivial background fields \cite{foc}. The Neveu--Schwarz sector
corresponds to the untwisted sector, and the sectors
$b_1$, $b_2$ and $b_3$ to the three twisted sectors
of the $Z_2\times Z_2$ orbifold model.
In addition to the gravity and gauge multiplets, the Neveu--Schwarz
sector produces six multiplets in the 10 representation
of $SO(10)$, and several $SO(10)$ singlets transforming under the
flavour $SO(6)^3$ symmetries.
The sectors $b_1$, $b_2$ and $b_3$ produce the Standard Model
matter fields that are embedded in the spinorial 16 representations
of $SO(10)$.
At the level of the NAHE set all the states in the free fermionic models
fall into representations of $SO(10)$, or are $SO(10)$ singlets. 
Furthermore, at this stage the hidden $E_8$ is unbroken, hidden matter
does not arise, and the models do not provide 
any candidates for the UHECR events. 

The second stage of the basis construction consists of adding
to the NAHE set three basis vectors, which
correspond to ``Wilson lines'' in the orbifold formulation.
The additional vectors reduce the number of generations
to three, one {}from each sector $b_1$, $b_2$ and $b_3$,
and break the gauge symmetries of the NAHE set.
The $SO(10)$ symmetry is broken to one of its subgroups
$SU(5)\times U(1)$ \cite{revamp}, $SO(6)\times SO(4)$ \cite{ALR},
$SU(3)\times U(1)_{B-L}\times SU(2)_L\times SU(2)_R$ \cite{lrs},
or $SU(3)\times SU(2)\times U(1)_{B-L}\times U(1)_{T_{3_R}}$ \cite{SLM}.
At the same time the hidden $E_8$ symmetry is broken to one of
its subgroups. 

The choice of the subgroup of $SO(10)$ that is left unbroken at
the string scale determines what kind of exotic matter states
can appear in a given string model, as well as affecting the final
subgroup of $E_8$ which remains
unbroken by the choice of boundary condition basis vector and GSO phases.
Since the superstring derived standard--like models contain
the symmetry breaking sectors that arise also in the other models,
their massless spectra admits also the exotic
representations that can appear in these models.
We therefore focus our discussion on the superstring standard--like models. 

In the superstring standard--like models,
the observable gauge group after application
of the generalized GSO projections is
$SU(3)_C\times U(1)_C\times SU(2)_L\times U(1)_L
\times U(1)^3\times U(1)^n$.
The electromagnetic charge is given by
\begin{equation}
U(1)_{\rm e.m.}=T_{3_L}+U(1)_Y,
\label{quem}
\end{equation}
where $T_{3_L}$ is the diagonal
generator of $SU(2)_L$, and $U(1)_Y$ is the weak hypercharge.
The weak hypercharge is given by{\footnote{
Note that we could have instead defined the weak hypercharge to be
$U(1)_Y={1\over 3} U(1)_C - {1\over 2} U(1)_L$. This amounts to the same
redefinition of fields between the straight and flipped $SU(5)$. In this
paper we will use the definition in Eq. \ref{qu1y}.}}
\begin{equation}
U(1)_Y={1\over 3} U(1)_C + {1\over 2} U(1)_L
\label{qu1y}
\end{equation}
and the orthogonal
combination is given by
\begin{equation}
U(1)_{Z^\prime}= U(1)_C - U(1)_L.
\label{quzp}
\end{equation}

The massless spectrum of the standard--like models contains
three chiral generations, each consisting of a 16 of
$SO(10)$, decomposed under the final $SO(10)$ subgroup as
\beqn
{e_L^c}&\equiv& ~[(1,~~{3\over2});(1,~1)]_{(~1~,~1/2~,~1)}~~;~~
{u_L^c}~\equiv ~[({\bar 3},-{1\over2});(1,-1)]_{(-2/3,1/2,-2/3)};~~~~
                                                        \label{ulc}\\
{d_L^c}&\equiv& ~[({\bar 3},-{1\over2});(1,~1)]_{(1/3,-3/2,1/3)}~~;~~
Q~\equiv ~[(3, {1\over2});(2,~0)]_{(1/6,1/2,(2/3,-1/3))};~~~~\label{q}\\
{N_L^c}&\equiv& ~[(1,~~{3\over2});(1,-1)]_{(~0~,~5/2~,~0)}~~;~~
L~\equiv ~[(1,-{3\over2});(2,~0)]_{(-1/2,-3/2,(0,1))},~~~~\label{l}
\eeqn
where we have used the notation
\begin{equation}
[(SU(3)_C\times U(1)_C);
     (SU(2)_L\times U(1)_L)]_{(Q_Y,Q_{Z^\prime},Q_{\rm e.m.})},
\label{notation}
\end{equation}
and have written the electric charge of the two
components for the doublets.

The matter states {}from the NS sector and the sectors $b_1$, $b_2$
and $b_3$ transform only under the observable gauge group.
In the realistic free fermionic models, there is typically one
additional sector that produces matter states transforming
only under the observable gauge group \cite{SLM}. These
states complete the representations that we identify with
possible representations of the Standard Model. In addition
to the Standard Model states, semi--realistic superstring
models may contain additional multiplets, in the $16$ and $\overline{16}$
representation of $SO(10)$, in the vectorial $10$ representation of $SO(10)$,
or the $27$ and $\overline{27}$ of $E_6$. Such states can pair up
to form super-massive states. They can mix with, and decay into,
the Standard Model representation unless some additional symmetry,
which forbids their decay, is imposed. For example, in the
flipped $SU(5)$ superstring models \cite{revamp}, two of the
additional vectors which extend the NAHE set produce an additional
$16$ and $\overline{16}$ representation of $SO(10)$. These states
are used in the flipped $SU(5)$ model to break the $SU(5)\times
U(1)$ symmetry to $SU(3)\times SU(2)\times U(1)$.

In addition to the states mentioned above transforming solely
under the observable gauge group, the sectors $b_j+2\gamma$
produce matter states that fall into the 16 representation of the hidden
$SO(16)$ gauge group decomposed under the final hidden gauge group.
The states {}from the sectors $b_j+2\gamma$ are $SO(10)$ singlets, but
are charged under the flavour $U(1)$ symmetries. All the states above
fit into standard representations of the grand unified group which may be,
for example, $SO(10)$ or $E_6$, or are singlets of these groups. They
carry the standard charges under the Standard Model gauge group or of
its GUT extensions. 

The superstring models contain additional states that cannot fit into
multiplets of the original $SO(10)$ unifying gauge group. They result {}from
the breaking of the $SO(10)$ gauge group at the string level via the
boundary condition assignment. The exotic states in the realistic free
fermionic models appear in vector--like representations and can acquire
a large mass.  Next we enumerate the exotic states that appear in free
fermionic models. The states are classified according to the unbroken
$SO(10)$ subgroup in each sector \cite{ccf}.

{}From the $\underline{SO(6)\times SO(4)}$ type sectors we obtain the
following exotic states.

\bigskip
$\bullet~ {\underline {\rm Colour ~triplets:}}~~~~
  [(    3, {1\over2});(1,0)]_{( 1/6, 1/2, 1/6)}~~~~;~~~~
  [({\bar3},-{1\over2});(1,0)]_{(-1/6,-1/2,-1/6)}$
\parindent=15pt
\bigskip

$\bullet~{\underline{\rm Electroweak ~doublets:}}~~~~
[(1,0);(2,0)]_{(0,0,\pm1/2)}$

\bigskip

$\bullet$ {\underline{Fractionally charged
                $SU(3)_C\times SU(2)_L$ singlets :}}
\beq
[(1,0);(1,\pm{1})]_{(\pm1/2,\mp1/2,\pm1/2)}~~~~;~~~~
[(1,\pm3/2);(1,0)]_{(\pm1/2,\pm1/2,\pm1/2)}
\label{fc64singlet}
\eeq
\parindent=15pt
The colour triplets bind with light quarks to form mesons and baryons
with  fractional electric charges $\pm1/2$ and $\pm3/2$.
The $\underline{SO(6)\times SO(4)}$ type states can appear 
in the Pati--Salam type models \cite{ALR} or in 
Standard--like models \cite{SLM}.

{}From sectors which break the $SO(10)$ symmetry into 
$\underline{SU(5)\times U(1)}$ we obtain exotic states
with fractional electric charge $\pm1/2$

\bigskip

$\bullet$ 
{\underline{Fractionally charged $SU(3)_C\times SU(2)_L$ singlets :}}
\beq
[(1,\pm3/4);(1,\pm{1/2})]_{(\pm1/2,\pm1/4,\pm1/2)}
\label{fc51singlet}
\eeq
\parindent=15pt

In general the fractionally charged states may transform
under a non--Abelian hidden gauge group in which case the fractionally
charged states may be confined.
For example, in the ``revamped'' flipped $SU(5)$ model \cite{revamp}
the states with fractional charge $\pm1/2$ transform as $4$ and $\bar4$
of the hidden $SU(4)$ gauge group. The states with the charges in
eq. (\ref{fc51singlet}) are called the ``cryptons'' and
may form good dark matter candidates \cite{eln}
if the lightest confined state is electrically neutral.
In the ``revamped'' flipped $SU(5)$ model it has been
argued that the lightest state is the ``tetron'',
which contains four fundamental constituents. 
In other models, states with the charges of eq. (\ref{fc51singlet})
may be singlets of all the non--Abelian group factors.

Finally in the superstring derived standard--like models
we may obtain exotic states {}from sectors which are combinations
of the $\underline{SO(6)\times SO(4)}$ breaking vectors and 
$\underline{SU(5)\times U(1)}$
breaking vectors. These states therefore arise only in
the $\underline{SU(3)\times SU(2)\times U(1)^2}$ type models. 
These states then carry the standard charges
under the Standard Model gauge group but carry fractional charges
under the $U(1)_{Z^\prime}$ gauge group.
The following exotic states are obtained:
\beq
\bullet~ {\underline{\rm colour ~triplets :}}~~~~
[(3,{1\over4});(1,{1\over2})]_{(-1/3,-1/4,-1/3)}~~~~;~~~~
[(\bar3,-{1\over4});(1,{1\over2})]_{(1/3,1/4,1/3)}
\label{unit}
\eeq

\bigskip
\parindent=15pt

Due to its potential role in string gauge coupling unification
\cite{gcu}, this state is referred to as ``the uniton'' \cite{ccf}.
The uniton forms bound states with the lightest up and down
quarks and gives rise to ultra--heavy mesons. In ref.~\cite{ccf}
it has been shown that the lightest meson can be the electrically
neutral state. 

\bigskip

$\bullet~ {\underline{\rm electroweak ~doublets : }}~~~~
[(1,\pm{3\over4});(2,\pm{1\over2})]_{(\pm1/2,\pm1/4,(1,0);(0,-1))}$

\bigskip
\parindent=15pt

Unlike the previous electroweak doublets, these electroweak doublets
carry the regular charges under the standard model gauge group but carry
``fractional'' charge under the $U(1)_{Z^\prime}$ symmetry. 
Finally, in the superstring derived standard--like models we also obtain
states which are Standard Model singlets but carry ``fractional''
charges under the $U(1)_{Z^\prime}$ symmetry.

\bigskip

$\bullet$ {\underline{Standard model singlets with ``fractional''
                        $U(1)_{Z^\prime}$ charge :}}

\beq
[(1,\pm{3\over4});(1,\mp{1\over2})]_{(0,\pm5/4,0)}
\label{fc321singlet}
\eeq
\parindent=15pt
These states may transform under a non--Abelian hidden gauge group
or may be singlets of all the non--Abelian group factors.
This type of Standard Model singlet appears in all the
known free fermionic standard--like models.
We refer to this state as the ``singleton''.

There are several important issues to examine
with regard to the exotic states. Since some of these states carry
fractional charges, it is desirable to make them sufficiently heavy
or sufficiently rare. A priori, in a generic string model, it is not
at all guaranteed that the states with fractional electric charge can
be decoupled or confined \cite{lykken}. 
Therefore, their presence imposes an highly
non--trivial constraint on potentially viable string vacua. 
In the NAHE--based free fermionic models, all the exotic
matter states appear in vector--like representations. They can therefore
obtain mass terms {}from renormalizable or higher order terms in the
superpotential. We must then study the renormalizable and
nonrenormalizable superpotential in the specific models.
The cubic level and higher order terms in the superpotential are
extracted by applying the rules of ref.~\cite{KLN}.
The problem of fractionally charged states
was investigated in ref.~\cite{fcp,cfn} for the model of ref.
\cite{fny}. 
By examining the fractionally charged states and the trilinear
superpotential, it was shown that all the fractionally charged
states receive a Planck scale mass by giving a VEV to
a set of $SO(10)$ singlets in the massless string spectrum. 
Therefore, all the fractionally charged states can decouple
{}from the remaining light spectrum. 
The second issue that must be examined with regard to the
exotic ``Wilsonian'' matter is the interactions with the Standard Model
states. The fractional charges of the exotic states
under the unbroken $U(1)$ generators of the $SO(10)$ gauge group,
may result in conserved discrete symmetry which forbid, or suppress,
their decay to the lighter Standard Model states \cite{ccf}.
  
In addition to the ``Wilsonian'' states the free fermionic 
models can give rise to semi--stable hidden glueballs that
arise {}from the unbroken subgroup of the hidden $E_8$ gauge 
group. Such states were considered as candidates for self--interacting
dark matter in ref.~\cite{maxim}. The hidden glueballs 
can interact with the Standard Model states only via
super-heavy hidden matter which is additionally charged
under the flavour $U(1)$ symmetries. 
Imposing that this
hidden glueballs accounts for the dark matter requires
that its self--interaction strength is of the order 
of hadronic interactions.

\section{Candidates}

\subsection{The fate of the cryptons}

The cryptons are fractionally charged states of the form 
of equation (\ref{fc51singlet}). This type of states 
appears often in realistic free fermionic models.
Experimental limits
on fractionally charged states impose that these states
either become sufficiently massive or are confined
into integrally charged states by some hidden sector
gauge group. In the flipped $SU(5)$ revamped model
the hidden gauge group is $SO(10)\times SU(4)$.
All the fractionally charged states in this model
transform under representations of the hidden $SO(10)$
or $SU(4)$ group factors. In table 1 we enumerate the spectrum
of hidden matter in the revamped flipped $SU(5)$ model. 

\begin{table}[t]
\begin{center}
\begin{tabular}{|l|l|}
\hline
$\Delta^0_1(0,1,6,0,-\frac{1}{2},\frac{1}{2},0)$&
$\Delta^0_2(0,1,6,-\frac{1}{2},0,\frac{1}{2},0)$\\[1ex]
$\Delta^0_3(0,1,6,-\frac{1}{2},-\frac{1}{2},0,\frac{1}{2})$&
$\Delta^0_4(0,1,6,0,-\frac{1}{2},\frac{1}{2},0)$\\[1ex]
$\Delta^0_5(0,1,6,\frac{1}{2},0,-\frac{1}{2},0)$& \\[1ex]
\hline 
 &
$T^0_1(10,1,0,-\frac{1}{2},\frac{1}{2},0)$ \\[1ex]
$T^0_2(10,1,-\frac{1}{2},0,\frac{1}{2},0)$ &
$T^0_3(10,1,-\frac{1}{2},-\frac{1}{2},0,\frac{1}{2})$ \\[1ex]
$T^0_4(10,1,0,\frac{1}{2},-\frac{1}{2},0)$& 
$T^0_5(10,1,-\frac{1}{2},0,\frac{1}{2},0)$  \\ [1ex]
\hline
\end{tabular}\\[1ex]
\begin{tabular}{|c|c|} \hline ${\tilde
F}^{+\frac{1}{2}}_1(1,4,-\frac{1}{4},\frac{1}{4},-\frac{1}{4},\frac{1}{2})$
& ${\tilde
F^{+\frac{1}{2}}}_2(1,4,-\frac{1}{4},\frac{1}{4},-\frac{1}{4},-\frac{1}{2})$
\\[1ex] ${\tilde
F}^{-\frac{1}{2}}_3(1,4,\frac{1}{4},\frac{1}{4},-\frac{1}{4},\frac{1}{2})$
& ${\tilde
F}^{+\frac{1}{2}}_4(1,4,\frac{1}{4},-\frac{1}{4},-\frac{1}{4},6-\frac{1}{2})$
\\[1ex] ${\tilde
F}^{+\frac{1}{2}}_5(1,4,-\frac{1}{4},\frac{3}{4},-\frac{1}{4},0)$ &
${\tilde
F}^{+\frac{1}{2}}_6(1,4,-\frac{1}{4},\frac{1}{4},-\frac{1}{4},
-\frac{1}{2})$\\[1ex] ${\tilde {\bar
F}}^{-\frac{1}{2}}_1(1,4,-\frac{1}{4},\frac{1}{4},\frac{1}{4},\frac{1}{2})$
& ${\tilde {\bar
F}}^{-\frac{1}{2}}_2(1,4,-\frac{1}{4},\frac{1}{4},\frac{1}{4},-\frac{1}{2})$
\\[1ex] ${\tilde {\bar
F}}^{+\frac{1}{2}}_3(1,4,-\frac{1}{4},-\frac{1}{4},\frac{1}{4},-\frac{1}{2})$
& ${\tilde {\bar
F}}^{-\frac{1}{2}}_4(1,4,-\frac{1}{4},\frac{1}{4},\frac{1}{4},
-\frac{1}{2})$ \\[1ex] ${\tilde {\bar
F}}^{-\frac{1}{2}}_5(1,4,-\frac{3}{4},\frac{1}{4},-\frac{1}{4},0)$ &
${\tilde {\bar
F}}^{-\frac{1}{2}}_6(1,4,\frac{1}{4},-\frac{1}{4},\frac{1}{4},
-\frac{1}{2})$\\[1ex] \hline
\end{tabular}
\end{center}
\caption{The spectrum of hidden matter fields 
that are massless at the string
scale in the revamped flipped $SU(5)$ model.
We display the quantum numbers under the hidden 
gauge group $SO(10) \times
SO(6) \times U(1)^4$, and subscripts indicate the electric charges.}
\end{table}

The states in the first table arise {}from the sectors $b_j+2\gamma$ and
$b_j+2\gamma+\zeta$. As noted above these states transform under vectorial
representations of the unbroken hidden $E_8$ subgroup, and are $SO(10)$
singlets. Consequently, they typically will obtain a mass term at a relatively
low order in the superpotential, and can have lower level interaction terms
in the superpotential. They affect the hidden sector dynamics
but do not give rise to fractionally charged exotics. The states
in the second table on the other hand, all transform as $4$ or
${\bar 4}$ of the hidden $SU(4)$ group factor, and carry fractional
electric charge $\pm1/2$. These states arise {}from the ``Wilsonian''
sectors, which are obtained {}from combinations of the $SO(10)$ breaking
basis vector with the other basis vectors that define the revamped 
flipped $SU(5)$ model. Analysis of the
renormalization-group $\beta$ functions of $SO(10)$ and $SO(6)$ suggests
that their confinement scales is of the order $\Lambda_{10}\sim
10^{14-15}$GeV for $SO(10)$ and  $\Lambda_{4}\sim 10^{11-12}$GeV for
$SU(4)$. This indicates that the $SU(4)$ states form the lightest
bound states.

The bound $SU(4)$ states are then composed of mesons, $T_iT_j$, 
$\Delta_i \Delta_j$ and ${\tilde  F_i} {\tilde {\bar F}_j}$,
baryons, $ {\tilde F_i} {\tilde F_j} \Delta_k$ and
${\tilde {\bar F}_i} {\tilde {\bar F}_j} \Delta_k$,
and quadrilinear {\it tetrons}, which are composed of 
four ${\tilde F}_i$s. The lightest of those have the forms
${\tilde F_i} {\tilde F_j}{\tilde F_k} {\tilde F_l}$ and
$ {\tilde {\bar F}_i}{\tilde {\bar F}_j}{\tilde {\bar F}_k}{\tilde
{\bar F}_l} $, where $i,j,k,l = 3,5$.
As in the case of QCD pions, one may expect the charged
states to be slightly heavier than the
neutral ones, due to electromagnetic energy mass
splitting. No non-renormalizable interaction capable of enabling
this lightest bound state to decay has been 
found in a search up to eighth order.
Generically, the crypton lifetime is expected to be given by
\beq
\tau_x\approx{1\over{m_x}}\left({M_S\over{m_x}}\right)^{2(N-3)},
\label{nthlifetime}
\eeq
where $m_x\sim\Lambda$ is the hidden sector confinement scale,
$M_S$ is the string scale and $N$ is order of the nonrenormalizable terms
that induce the tetron decay. Taking $N=8$, $M_S\sim10^{17-18}\;{\rm GeV}$
and a tetron mass $m_x\sim10^{12}\;{\rm GeV}$, one finds that
$\tau_x>10^{7-17}\;{\rm years}$.
It has therefore been suggested that the lightest neutral tetron
is a perfect candidate for a superheavy dark matter particle. 

However, as discussed above, in addition to the lightest 
neutral tetron charged tetrons are formed as well. These can be argued
to be slightly heavier than the neutral tetron due to the electromagnetic
splitting. In order not to be over--abundant today
the charged tetrons should decay into the neutral tetron.
However, the charged tetron is composed of the same 
constituents that compose the neutral tetron. Namely,
it is composed of the constituent cryptons that carry fractional
electric charge $\pm1/2$. Suppose then that the charged tetron is
composed of three $+1/2$ charged cryptons and one $-1/2$ charged 
crypton. In order to get a neutral tetron one of the $+1/2$ cryptons
has to convert to a $-1/2$ crypton. However, since the cryptons are 
$SU(2)_L$ singlets, they cannot convert by emitting a charged $W^+$
gauge boson, or a charged Higgs. That is, the charged tetron
cannot convert to a neutral one by emitting a light degree of freedom.  
Therefore, the only way for the charged tetron to convert to a neutral
one is by exchanging a charge $+1$ heavy degree of freedom. The only 
possible such states are the heavy charged gauge bosons or massive string
states. Effectively, therefore, the only way for the charged tetron
to decay to the neutral one is by the same higher order nonrenormalizable
operators which govern the decay of the neutral tetrons. In effect
the reason is that both the neutral and charged tetron decay arise 
{}from annihilation, through the higher order nonrenormalizable terms,
of the constituent cryptons inside the tetrons. 
The conclusion 
is therefore that if the neutral tetron is long lived, so will be the
charged tetron. The abundance of the charged tetron is therefore 
comparable to the abundance of the neutral tetron. As stable charged
matter is strongly constrained,
this argument therefore indicates that the tetron cannot provide
an appealing candidate for cold dark matter. 
One should of course qualify this statement by admitting that it is
of course not impossible that specific models will contrive 
to give rise to operators which allow charged tetron decay while they
forbid the neutral tetron decay. However, we note that in the
revamped $SU(5)$ this is not the case as such operators do not arise 
up to order $N=8$ nonrenormalizable terms. 

To conclude this discussion, we note that the crypton passes two 
of the criteria that are needed to produce an appealing superheavy
dark matter candidate, while it fails on the third. Namely, its
mass scale is generated by the hidden sector strong dynamics and
its stability arises {}from the ``Wilsonian'' symmetry breaking mechanism
and the resulting fractional electric charge. However, the fact that
it carries fractional electric charge implies that also the 
charged tetrons are long lived and give rise to the stable charged
matter which is severely constrained. It is still possible, however,
that the tetron exists in sufficient abundance to account for
the UHECR events, while it is sufficiently diluted to evade
the charged dark matter constraints. Next we examine this possibility.

\subsection{Abundance of charged and neutral tetrons \label{abundances}}

Tetrons cannot have been in thermal equilibrium in the early
universe, since then their freeze-out energy density $\rho_T$
would be larger than the critical density $\rho_c$ of the
universe \cite{Griest}. Hence, they must have been produced
non-thermally. It has been shown that particles with
masses $\gsim10^{12}\;$GeV can efficiently be produced
gravitationally \cite{Kolb3}, even if the reheating temperature
is much lower than their mass. Since charged and
neutral tetrons have almost identical masses, they should be
produced in equal abundances in gravitational production
mechanisms. On the other hand, heavy particles can also
be produced non-thermally during preheating \cite{Kolb1}
or reheating \cite{Kolb2}. Then the electromagnetic interactions
of charged tetrons might even lead to a charged tetron abundance
larger than the neutral tetron abundance, if the freeze-out
temperature of charged tetrons is smaller than the maximum
temperature reached during reheating. Hence, the charged
tetron abundance is going to be at least of the same order
as the neutral tetron abundance.

However, there are strong bounds on the abundance of long-lived
charged massive particles (CHAMPs) \cite{Perl}. Indeed, if neutral
and charged tetrons are to constitute cold dark matter (CDM),
one would expect a certain flux of CHAMPs which should be
measurable, e.g.\ in GUT monopole detectors. Results {}from
MACRO \cite{Macro} and a surface scintillator array \cite{Barish}
place bounds on such a flux which are well below the expected
dark matter flux \cite{Perl}. Further, charged tetrons are
captured in disk stars and can destroy neutron stars on a time
scales $\lsim10\;$years, i.e.\ for a mass $\sim10^{12}\;$GeV
the tetron energy density can be at most \cite{Gould}
\begin{equation}
 \Omega_T\lsim 10^{-6} \Omega_{CDM}\;.
 \label{t_abund}
\end{equation}
Hence, tetrons cannot be cold dark matter. However, they still
can be responsible for the ultrahigh energy cosmic rays if
\cite{Berezinsky}:
\begin{equation}
 {\Omega_T\over\Omega_{CDM}}{t_U\over\tau_T}\sim5\times10^{-11}\;,
 \label{lifetime}
\end{equation}
where $t_U$ is the age of the universe and $\tau_T$ the lifetime
of these tetrons. Hence, the upper bound (\ref{t_abund}) yields
an upper bound on the tetron lifetime $\tau_T\lsim 2\times10^4t_U$.

\subsection{Renormalization group analysis}

In the previous subsections we showed that the crypton's
abundance is severely limited by constraints on stable heavy charged matter,
but it may still be sufficiently longed--lived and abundant
to account for the UHECR events.
Additionally, the cryptons affect the renormalization
group equations of the Standard Model parameters. 
In this section we study the effect on gauge coupling unification.

In the revamped flipped $SU(5)$ model the cryptons carry fractional
charge $\pm1/2$ and transform as $4$ and $\bar4$ under the hidden
$SU(4)$ gauge group. As discussed in refs. \cite{eln,ben}, {}from analysis
of the superpotential up to order $N=6$ it is noted that the states 
${\tilde {F}}^{+\frac{1}{2}}_{3,5}$ and
${\tilde {\bar F}}^{+\frac{1}{2}}_{3,5}$ remain massless and confine 
to form the bound tetron states. These states therefore contribute to
the evolution of the Renormalization Group Equations (RGEs) of the 
Standard Model gauge parameters. In the perturbative heterotic--string the 
Standard Model gauge couplings are unified at the string scale, which is
of the order \cite{Kaplunovsky}
\beq
    M_S~\equiv~M_{\rm string}~\approx~g_{\rm
      string}\,\times\,5\,\times\,10^{17} ~\;{\rm GeV}~,
\label{hetstringscale}
\eeq
where $g_{\rm string}$ is the unified string coupling.
The one--loop RGEs for the Standard Model
gauge couplings are given by, 
\beq
     {{16\pi^2}\over{g_i^2(\mu)}}~=~k_i{{16\pi^2}\over{g_{\rm string}^2}}~+~
     b_i\,\ln{ M^2_{\rm string}\over\mu^2}~+~\Delta_i^{\rm (total)}
\label{reminder}
\eeq
where $b_i$ are the one-loop beta-function coefficients, and
the $\Delta_i^{\rm (total)}$
represent possible corrections {}from the additional gauge or matter states.
By solving (\ref{reminder}) for $i=1,2,3$ simultaneously,
we obtain expressions for $\sin^2\theta_W(M_Z)$ and
$\alpha_{3}(M_Z)$, which are confronted with the experimentally
measured values for these observables at the $Z$--boson mass scale.
The expression for $\alpha_{3}(M_Z)$
then takes the general form
\beqn
    \alpha^{-1}_{3}(M_Z)\vert_{\overline{MS}}&=&
      \Delta_{\rm MSSM}^{(\alpha)} +
      \Delta_{\rm h.s.}^{(\alpha)} +
      \Delta_{\rm l.s.}^{(\alpha)} +
       \Delta_{\rm i.g.}^{(\alpha)} +\nonumber\\
      &&~~~\Delta_{\rm i.m.}^{(\alpha)} +
      \Delta_{\rm 2-loop}^{(\alpha)} +
      \Delta_{\rm Yuk.}^{(\alpha)} +
      \Delta_{\rm conv.}^{(\alpha)},
\label{generalform}
\eeqn
and likewise for $\sin^2\theta_W(M_Z)\vert_{\overline{MS}}$ with
corresponding corrections $\Delta^{\rm (sin)}$.
Here $\Delta_{\rm MSSM}$ represents the one-loop
contributions {}from the spectrum of the Minimal Supersymmetric
Standard Model (MSSM) between the unification scale and the $Z$ scale, and
the remaining
$\Delta$ terms respectively correspond
to the second-loop corrections, the Yukawa-coupling corrections,
the corrections {}from scheme conversion,
the heavy string thresholds, possible light SUSY threshold corrections,
corrections {}from possible additional intermediate-scale gauge structure
between the unification and $Z$ scales,
and corrections {}from possible extra intermediate-scale matter.
Each of these $\Delta$ terms has an algebraic
expression in terms of $\alpha_{\rm e.m.}$ as well
as model-specific parameters such as $k_1$,
the beta-function coefficients, and the
appropriate intermediate mass scales.
Here we take $k_1=5/3$, which is the canonical $SO(10)$ value.
We also neglect here the small effect \cite{gcu}
of $\Delta_{\rm h.s.}^{(\alpha)}$, $\Delta_{\rm l.s.}^{(\alpha)}$,
$\Delta_{\rm 2-loop}^{(\alpha)}$, $\Delta_{\rm Yuk.}^{(\alpha)}$
$\Delta_{\rm conv.}^{(\alpha)}$, as 
we are only interested in the qualitative effect of the intermediate
crypton matter. 

\begin{figure}[t]
\centerline{\epsfxsize 3.0 truein \epsfbox {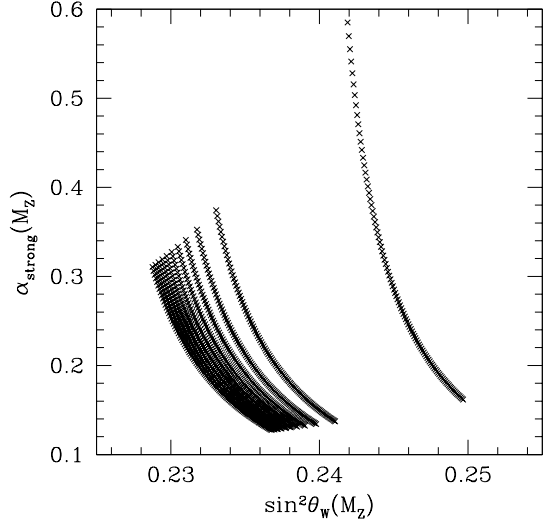}}
\caption{Scatter plot of $\sin^2\theta_W(M_Z)$
versus $\alpha_s(M_Z)$ for the entire range in eq. (\ref{mcms}).}
\label{s2was}
\end{figure}
\begin{figure}[tbh]
\centerline{\epsfxsize 3.0 truein \epsfbox {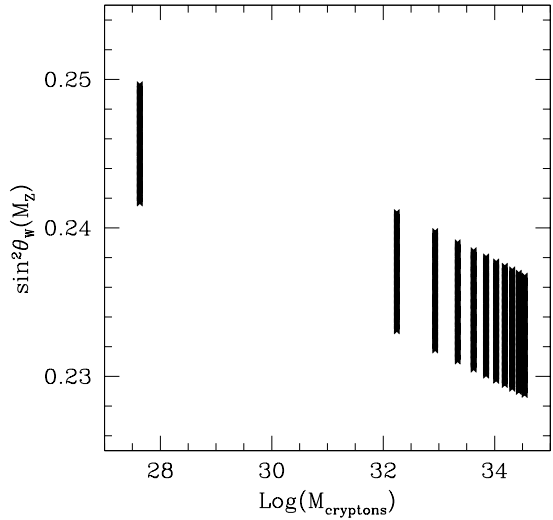}}
\caption{Scatter plot of $\sin^2\theta_W(M_Z)$
versus ${\rm Log}(M_{\rm crypton})\equiv\Lambda_4$
for the range in eq. (\ref{mcms}).}
\label{s2wmc}
\end{figure}
\begin{figure}[tbh]
\centerline{\epsfxsize 3.0 truein \epsfbox {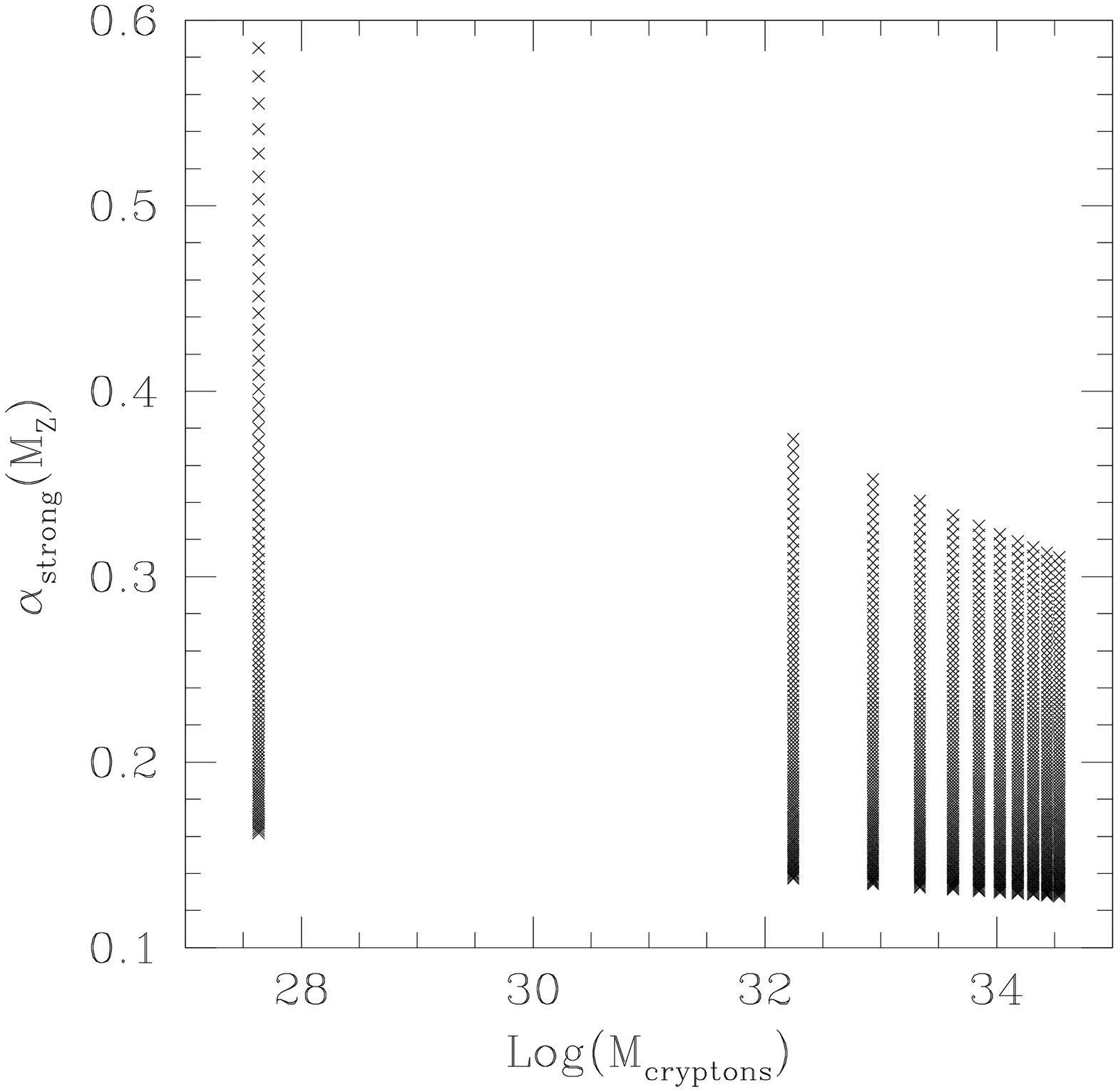}}
\caption{Scatter plot of $\alpha_s(M_Z)$
versus ${\rm Log}(M_{\rm crypton})\equiv\Lambda_4$
for the range in eq. (\ref{mcms}).}
\label{asmc}
\end{figure}

For $\sin^2\theta_W (M_Z)$ we have:
\beqn
    \Delta^{(\sin)}_{\rm MSSM} &=&
         {1\over{k_1+1}}\left\lbrack
     1-{a\over{2\pi}}(11-k_1)\ln{M_S\over{M_Z}}\right\rbrack
                \nonumber\\
    \Delta^{(\sin)}_{\rm i.m.} &=&
      {1\over{2\pi}}\,\sum_{i}\,{{k_1a}\over{k_1+1}}\,
             (b_{2_i}-b_{1_i})\ln{M_S\over{M_i}}
\label{s2wmz}
\eeqn
where $M_S\equiv M_{\rm string}$ is the string unification scale,
$a\equiv \alpha_{\rm e.m.}(M_Z)$,
$M_i$ are the intermediate matter mass scales.
Likewise, for $\alpha^{-1}_{3}(M_Z)$, we have:
\beqn
    \Delta^{(\alpha)}_{\rm MSSM} &=&
      {1\over{1+k_1}}\,\left\lbrack {1\over a}+{1\over{2\pi}}(-3 k_1-15)
       \ln{M_S\over{M_Z}}\right\rbrack\nonumber\\
    \Delta^{(\alpha)}_{\rm i.m.} &=&
        -~{1\over{2\pi}}\,\sum_{i}\,\left\lbrack
\left({k_1\over{1+k_1}}\right)b_{1_{i}}+
       \left({1\over{1+k_1}}\right)b_{2_{i}}
 -b_{3_{i}}\right\rbrack\ln{M_{S}\over{M_i}}
\label{a3invmz}
\eeqn

An important issue in string unification is that of the unification
scale. The perturbative heterotic string predicts the scale of eq. 
(\ref{hetstringscale}). In this case string gauge coupling unification
necessitates the existence of additional $SU(3)_{\rm colour}$ matter
representations \cite{gcu}. However, nonperturbative string dualities
reveal that the nonperturbative string unification can be lower
\cite{witten}, and can be compatible with the MSSM unification scale 
\cite{mssmunification} 
without the need for additional colour matter states. We discuss
the consequences of crypton matter in both cases.

The beta--function coefficients for the cryptons ${\tilde F}$ and
$\overline{\tilde F}$ are
\beq
      \pmatrix{ b_{SU(3)} \cr b_{SU(2)} \cr b_{U(1)}
\cr}_{{\tilde F},\overline{\tilde F}}~=~
             \pmatrix{0\cr 0\cr 3/5\cr}~.
\label{crypbetafun}
\eeq
Since we have a total of four massless representations at the scale
$\Lambda_4$, the total contribution of the 
crypton matter states above that scale to the $U(1)$ beta--function is 12/5.
We insert these numbers into the equations for the predicted values
of $\alpha_3^{-1}(M_Z)$ and $\sin^2\theta_W(M_Z)$, 
and vary
\beqn
10^{12}\;{\rm GeV}\le & \Lambda_4 ~\le 10^{15}\;{\rm GeV} \nonumber\\
2\cdot 10^{16} \;{\rm GeV}\le  & ~{M}_{S}~~ 
\le 5\cdot10^{17}\;{\rm GeV}. \label{mcms}
\eeqn
In this we in effect assume the spectrum of the MSSM plus the
additional crypton states, and ignore the possible effect of the
intermediate GUT threshold. The variation of the string unification
scale incorporates the possible non--perturbative string effects,
while that of the crypton mass scale incorporates the possible variation 
of the meta--stable crypton mass scale.  
In fig (\ref{s2was}) we present a scatter plot of $\sin^2\theta_W(M_Z)$
versus $\alpha_s(M_Z)$ for the entire range in eq. (\ref{mcms}). 
In figs. (\ref{s2wmc}) and (\ref{asmc}) we plot $\sin^2\theta_W(M_Z)$
and  $\alpha_s(M_Z)$, respectively, versus
${\rm Log}(M_{\rm crypton})\equiv\Lambda_4$.
{}From fig. (\ref{s2was}) we note that for the entire range in eq. (\ref{mcms})
viable values for $\sin^2\theta_W(M_Z)$
and  $\alpha_s(M_Z)$ are not obtained. 
This is of course somewhat expected as the additional matter affects 
these observables, which are in agreement with experiments
if one assumes solely the MSSM spectrum in the desert. 
There may exist, however, a priori an interplay between the effect 
of additional matter and scale variation, which can result in an
agreement with the experimental data. In fig. (\ref{s2wmc})
we see that lower $\Lambda_4$ results in stronger disagreement with
the data, whereas for $\Lambda_4\sim10^{15}\;{\rm GeV}$ viable
values for $\sin^2\theta_W(M_Z)$ can be obtained. Similar dependence
is noted in fig. (\ref{asmc}) for $\alpha_s(M_Z)$, with
$\alpha_s(M_Z)\ge0.127$.

In the case of the flipped $SU(5)$ model,
the effect of the intermediate scale can be incorporated
via the following additional term in (\ref{s2wmz}):
\beq
        -{1\over 2\pi}\, {32\over 5}\, {k_1 a\over k_1+1}\,
         \ln\,{M_S\over M_I}~.
\eeq
Here we have assumed the spectrum below $M_I$ to be that of
the MSSM, and above $M_I$ to consist of three {\bf 16} representations
of $SO(10)$, one ${\bf 5}$ and $\overline{\bf 5}$
of $SU(5)$ that produces the light Higgs doublets,
and one {\bf 10} and $\overline{\bf 10}$ of $SU(5)$
that is used to break the $SU(5)\times U(1)$ symmetry
to $SU(3)\times SU(2)\times U(1)$. In addition to the variation in eq. 
(\ref{mcms}), we vary the $SU(5)\times U(1)$ breaking scale 
between $M_x\equiv 2\cdot 10^{16}\;{\rm GeV}$ and $M_s$ as 
given in eq. (\ref{mcms}). 

\begin{figure}[t]
\centerline{\epsfxsize 3.0 truein \epsfbox {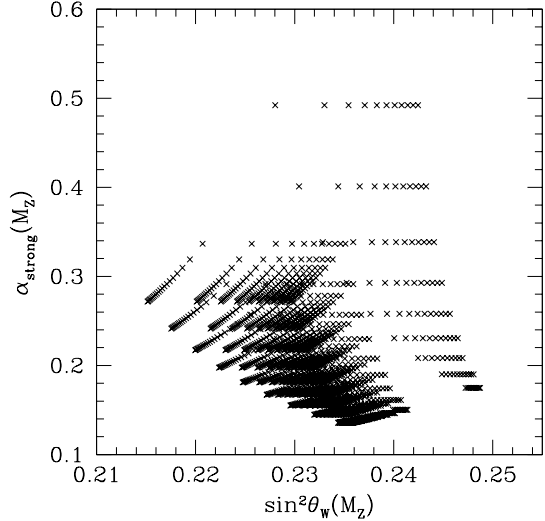}}
\caption{Scatter plot of $\sin^2\theta_W(M_Z)$
versus $\alpha_s(M_Z)$ for the entire range in eq. (\ref{mcms})
in the presence of a $SU(5)\times U(1)$ GUT threshold.}
\label{s2wfs5}
\end{figure}

\begin{figure}[tbh]
\centerline{\epsfxsize 3.0 truein \epsfbox {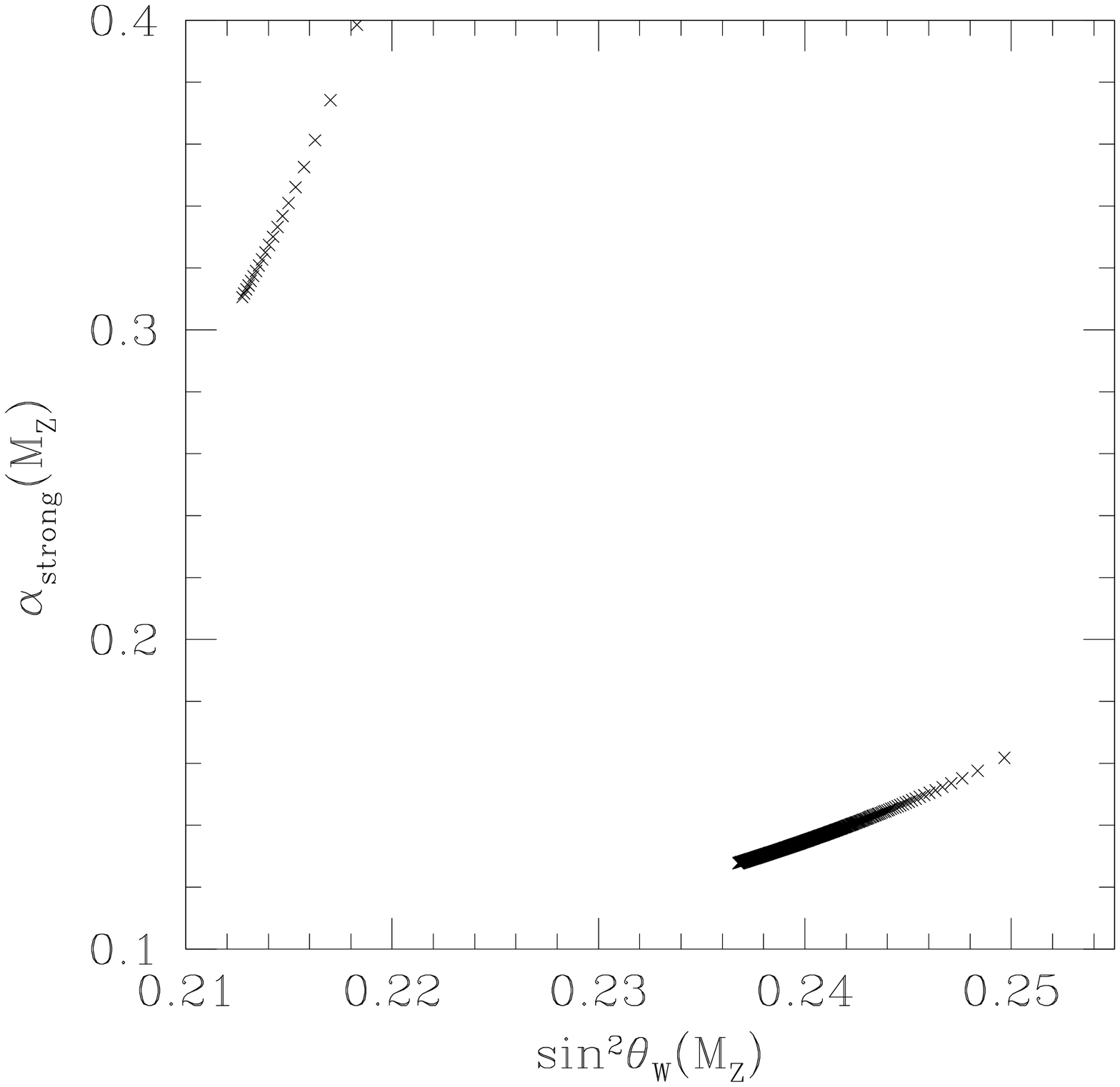}}
\caption{Scatter plot of $\sin^2\theta_W(M_Z)$
versus $\alpha_s(M_Z)$ for fixed values of $M_S$ and $M_5$,
and $\Lambda_4$ varied as in eq. (\ref{mcms}).}
\label{fmsm5s2was}
\end{figure}

The existence of the intermediate $SU(5)\times U(1)$ GUT threshold 
does not affect the prediction for $\alpha_3^{-1}(M_Z)$, as compared
to the extrapolation in the absence of this threshold. The reason
being that the evolution of $\alpha_3$ and $\alpha_2$ is identical
above this threshold. Hence, the dependence on their running above this 
threshold drops out when their evolution equations are equated at the
unification  scale. 
The effect of the extended gauge structure in this model
is to reduce $\sin^2\theta_W(M_Z)$.
In fig. (\ref{s2wfs5}) we present a scatter plot of 
$\sin^2\theta_W(M_Z)$
versus $\alpha_s(M_Z)$ for the entire range in eq. (\ref{mcms}),
including the variation of the intermediate $SU(5)\times U(1)$ 
GUT breaking threshold. Again we find that experimentally viable
values are not obtained. 
In fig. (\ref{fmsm5s2was}) we show a plot of $\sin^2\theta_W(M_Z)$
versus $\alpha_s(M_Z)$ for a fixed value of $M_5=2\cdot10^{16}\;{\rm GeV}$
and two fixed values of $(M_S=M_5;M_S=5\cdot10^{17}\;{\rm GeV}$,
and the crypton mass scale is varied as in eq. (\ref{mcms}).
The sparse (denser) curves correspond to the high (low) $M_S$ scales,
respectively. As expected the high $M_S$ scale results in stronger
disagreement with $\alpha_s(M_Z)$. However, also for the low 
$M_S$ scale we see that  $\sin^2\theta_W(M_Z)\ge0.237$,
and, as expected, approaches the experimentally allowed region
for the larger values of $\Lambda_4$. 
The conclusion of this analysis is that the presence of the 
crypton matter states at the intermediate scale, and 
agreement with the experimental observables 
$\sin^2\theta_W(M_Z)$ and $\alpha_s(M_Z)$
necessitates the addition of additional matter beyond the MSSM.
This constraint will of course be evaded if the 
exotic matter states are entirely neutral with respect
to the Standard Model gauge group.

\subsection{The destiny of the uniton}

The arguments above prompts us to examine whether other viable candidates
for CDM and the UHECR events can arise {}from string models. We study
the cases of the  uniton, the standard model singlet and the hidden sector 
glueballs.  
In this subsection we examine whether the uniton can provide a
viable candidate for the UHECR events.

The quantum numbers of the 
uniton are given in eq. (\ref{unit}). The uniton is a strongly
interacting particle and carries the regular down--quark
type charges under the Standard Model gauge group. 
It forms bound states with the light up and down quarks. 
The uniton carries
fractional charge under the $U(1)_{Z^\prime}$ which is 
embedded in $SO(10)$ and is orthogonal to the weak--hypercharge.
The fractional $U(1)_{Z^\prime}$ charge can result in local discrete
symmetries that suppress its decay to the Standard Model states
\cite{ccf,lds}. While this is model dependent here we assume that this
is indeed the case. In \cite{ccf} it was argued that the uniton
can give rise to a viable dark matter candidate, provided that
the neutral meson is lighter than the charged one. In \cite{fop}
it was shown that because of its strongly interacting light degree
of freedom the uniton gets trapped in the sun and the earth in a
substantial rate and subsequently annihilates into quarks and leptons.
This constrains the uniton mass to be above $10^{11}\;$GeV, in perfect 
agreement with the energy scale suggested by the UHECR events.
The decay of the bound uniton state can then arise due to
nonrenormalizable operators, yielding a lifetime similar
to eq. (\ref{nthlifetime}).

The pertained semi--stability of the uniton arises {}from its
fractional charge under $U(1)_{Z^\prime}$. The neutral meson
is assumed to be lighter than the charged one, in which case
the uniton might provide a viable dark matter candidate.
Like in the case of the crypton
the question then is whether the charged meson can decay into the 
neutral one at a sufficient rate. The uniton itself is
an $SU(2)_L$ singlet and therefore, similar to the crypton, cannot
emit a light $W^\pm$ gauge boson. However, the heavy bound charged
meson, contrary to the case of the tetron, is composed of the exotic
uniton and a Standard Model up--type quark, which can decay
through weak interactions.
Therefore, the charged uniton bound
state can decay into the neutral one by a Standard Model beta-decay.
Hence, the uniton is a viable candidate for cold dark matter
and UHECRs, as long as the neutral bound state is lighter than
the charged one.

In regard to the mass scale of the uniton, unlike the crypton 
it does not transform under the hidden non--Abelian gauge sector. 
Therefore, its mass scale is not fixed by the hidden sector dynamics.
The limits {}from trapping in the sun and earth, however, do constrain it to be 
in the region which is interesting {}from the perspective of the UHECR 
events. However, the mass scale is not generated dynamically and in
a sense has to be put in by hand. 

As has been discussed in section (\ref{abundances}), meta--stable relics 
with masses $>10^{11}\;$GeV would overclose the universe if they had been
in thermal equilibrium in the early universe. Hence, we have to consider
non-thermal production mechanisms. It has been shown that super-heavy 
particles can efficiently be produced gravitationally \cite{Kolb3}, or
during preheating \cite{Kolb1} or reheating \cite{Kolb2} after 
inflation. If we assume that meta--stable unitons form cold dark 
matter, their decays can explain the ultrahigh energy cosmic rays
if their lifetime is of the order (cf.~eq.~(\ref{lifetime}))
\begin{equation}
  \tau_{uniton}\sim 2\cdot10^{10}t_U\;,
\end{equation}
where $t_U$ is the age of the universe.

To summarize, the proclaimed stability of the uniton arises {}from
the ``Wilsonian'' gauge symmetry breaking. Unlike the case of the 
crypton the required mass scale does not originate {}from hidden sector
dynamics. As the uniton appears in vector--like representations,
the required mass scale can be generated by a direct superpotential
mass term. Such mass terms, of the required order of magnitude
can arise {}from nonrenormalizable terms. In this respect it is
appealing that the phenomenological constraints on uniton dark matter
impose $m_{\rm uniton}>10^{11}\;{\rm GeV}$ \cite{fop}. Lastly, the uniton
is a strongly interacting particle, and one has to address experimental
constraints on strongly interacting massive particles (SIMPs) as cold
dark matter. Unitons will interact with ordinary matter and the cross
section for their interactions with ordinary matter has been estimated
to be \cite{fop}
\begin{equation}
 \sigma\sim10^{-26}\;\mbox{cm}^2\;.
\end{equation}
Experimental constraints on SIMPs have recently been reanalyzed 
\cite{McGuire}, and it has been shown that there are several unconstrained
windows in the dark matter-proton cross section versus mass parameter
space. In particular, SIMPs with masses $>10^{11}\;$GeV and cross sections
with ordinary matter $<10^{-22}\;\mbox{cm}^2$ are viable cold dark matter
candidates.

We further comment that it is of course true that the
uniton also affects the evolution of the Standard Model
gauge couplings. However, the original motivation to introduce
the uniton at an intermediate energy scale was to enable heterotic
string gauge coupling unification. Hence, the uniton can help
to solve several problems at once,  by enabling heterotic--string 
unification, by providing a substantial fraction of the dark matter, 
and by explaining the observed UHECR events.

\subsection{The role of the singleton} 

The next candidate that we study is the Standard Model singlet, eq. 
(\ref{fc321singlet}), which carries fractional $U(1)_{Z^\prime}$
charge. We refer to this state as the ``singleton'', and it also
arises {}from the Wilson line breaking of the $SO(10)$ gauge symmetry.
Therefore, such states can be meta--stable due to the fractional 
$U(1)_{Z^\prime}$ charge, which may leave a
residual local discrete symmetry after the $U(1)_{Z^\prime}$ gauge
symmetry breaking. For example, this will be the case if the states
that break the $U(1)_{Z^\prime}$ gauge symmetry carry only ``integral''
$U(1)_{Z^\prime}$ charges\footnote{By ``integral'' here we mean charges
that are compatible with $SO(10)$ embedding.}.
In the model of ref.~\cite{eu} some of the singleton states transform
under the $SU(5)$ or $SU(3)$ hidden gauge groups, whereas others are singlets
of all the non--Abelian groups in the model. One may then envision a scenario
in which, after cancellation of the anomalous $U(1)$ $D$--term,
the remaining light states are those that transform under a hidden
non--Abelian gauge group. Then in a similar fashion to the generation of
the crypton mass, the mass scale of the singleton can arise {}from the
hidden sector dynamics. Finally the singleton is a Standard Model
singlet and therefore there is no danger of producing stable
charged relics. On the other hand, if we assume that the
singleton transforms under a hidden gauge group, say $SU(4)$,
then, similarly to the cryptons, it will form semi--stable bound 
states whose decay is similarly governed by the higher order 
nonrenormalizable operators, which do not involve $U(1)_{Z^\prime}$
breaking VEVs, and do not violate the assumed local discrete
symmetry that protect the singleton {}from decaying. Then, the 
meta--stability of the bound state arises {}from annihilation
of the constituent singletons which is induced by the nonrenormalizable
terms. We therefore conclude that the singleton states can provide 
appealing candidates for the UHECR events that satisfy the three desired 
criteria. Namely, their meta--stability originates {}from the Wilson line
breaking effect; their mass scale may originate {}from hidden sector 
dynamics and perhaps most importantly, contrary to the crypton or the uniton,
they are Standard Model singlets and therefore are electrically neutral. 
Finally, the non-thermal production mechanisms discussed in section
(\ref{abundances}) can give rise to a non-negligible singleton density,
i.e., singletons are attractive cold dark matter candidates and
can solve the ultrahigh energy cosmic ray problem.

\subsection{Other candidates}

The last candidate that we discuss are the exotic glueballs
of reference \cite{maxim}. These states were proposed there as 
candidates for strongly interacting dark matter. The basic
idea is that the hidden $E_8$ gauge group of the heterotic 
string models is typically broken to a $SU(n)\times SU(m)\times U(1)^k$
subgroup, with matter states in vector--like representations.
The matter states obtain intermediate scale mass terms, 
and the $SU(n)$ group factors condense at some scale below
that, depending on the matter and gauge content. At that scale 
exotic glueballs form, the lightest of which is expected to
be stable. The exotic glueballs can interact with the Standard Model
states only through the heavy matter states, which are charged
also with respect to the horizontal $U(1)$ symmetries. The flavour
symmetries are broken near the string scale and therefore
the exotic glueballs can interact with the Standard Model states
through higher order operators, which are suppressed by the
heavy mass scale. In ref.~\cite{spergel}
it was argued that the self--interacting
dark matter should have self--interactions of the order of the 
hadronic scale. This requirement therefore suggests that
the exotic glueballs could not provide suitable candidates
for the UHECR events.

\section{Conclusions}

The discovery of ultrahigh energy cosmic ray events
with energies exceeding the GZK bound is one of the most
intriguing recent experimental observations.
Semi--stable superheavy matter states provide
a plausible explanation for these events. 
In this paper we studied whether such a state could arise
from string theory. Heterotic string theory gives rise
to fractionally charged meta--stable Wilsonian matter states.
Thus, the heterotic string,
while allowing the embedding of the standard GUT structures,
at the same time also produces meta--stable massive states. 
We discussed
the various Wilsonian matter states that arise in the 
string models and classified them according to their
charges under the unbroken $SO(10)$ subgroup at the string scale.
We argued that states that carry fractional electric
charge, which are confined by a hidden sector gauge group, 
produce semi--stable charged matter in addition to the
lightest semi--stable neutral state. They are therefore 
severely limited by constraints on charged dark matter,
and cannot give rise to a viable dark matter candidate.
However, they may still be responsible for the observed
UHECR events, if their lifetime is of the right order
of magnitude.
Further, fractionally
charged matter at intermediate energy scales also 
affects the evolution of the Standard Model
parameters and we showed that it would necessitate
additional colour triplets and electroweak
doublets to enable the gauge couplings to unify. 
While this does not exclude the cryptons as candidates for 
the UHECR events, it makes them, in our opinion, less
attractive.
We further showed
that other Wilsonian matter states that carry the standard 
charges with respect to the Standard Model gauge group
but carry fractional charges with respect to the $SO(10)$
$U(1)_{Z^\prime}$ generator, can give rise to viable dark matter
candidates that can also account for the UHECR events. 
The meta--stability of such states arises due to the
Wilson line symmetry breaking mechanism and the fact that
the Standard Model states are obtained from representations
of the underlying GUT symmetry group. On the other hand
the decay of such states would similarly arise from 
higher order nonrenormalizable operators. 
The other properties of these states render them more, or less,
attractive.
While in the
case of the singleton the required mass scale could arise from hidden
sector dynamics, in the case of the uniton it has to be put in
by hand. However, it is very intriguing that the uniton
is in fact constrained to be heavier than $10^{11}\;{\rm GeV}$
\cite{fop}, which is in perfect harmony with the 
mass scale required to explain the UHECR events.  
Lastly, additional coloured matter states
at intermediate energy scale have been motivated from 
heterotic string gauge coupling unification. 
The singleton on the other hand provides the most likely
dark matter candidate in the sense that it is
a Standard Model singlet and therefore does not give rise
to charged or coloured matter. Finally, forthcoming
experimental data \cite{auger} and improved
theoretical analysis along the lines of ref.~\cite{fragfun}
promise exciting new results with potentially ground--breaking
discoveries.
\vspace{1cm}

\centerline{\bf Acknowledgements}
\smallskip
We thank Graham Ross, Subir Sarkar and Ramon Toldra for
useful discussions. AF would like to thank the Theory
groups at CERN and Lecce for hospitality. 
The work of CC is supported in part by INFN 
(iniziativa specifica BARI-21) and by MURST. The work
of AF is supported by PPARC, and M.P.\ was supported
by the EU network ``Supersymmetry and the early universe''
under contract no.~HPRN-CT-2000-00152.

\newpage

\end{document}